\documentstyle[12pt]{article}
\newcommand{\sbf}[1]{\mbox{{\scriptsize$\bf{#1}$}}}

\def\comment#1{}
\def\lfrac#1#2{#1/#2}

\font\twlmbf=cmmib10 scaled1200
\font\egtmbf=cmmib10 scaled800

\input{tcilatex}
\begin{document}

\title{Relativistic Trace Formula for Bound States in Terms of Classical Periodic
Orbits}
\author{H. Kleinert\thanks{kleinert@physik.fu-berlin.de ,
~http://www.physik.fu-berlin.de/\~{}kleinert \hfil
} \\
Institut f\"ur Theoretische Physik\\
Freie Universit\"at Berlin, Arnimallee 14, 1000 Berlin 33, Germany \and D.
H. Lin\thanks{%
e-mail: d793314@phys.nthu.edu.tw} \\
Department of Physics, National Tsing Hua University \\
Hsinchu 30043, Taiwan, Republic of China\\
}
\maketitle

\setlength{\baselineskip}{1cm} \centerline{\bf Abstract} {We set up a trace
formula for the relativistic density of states in terms of a topological sum
of classical periodic orbits. The result is applicable to any relativistic
integrable system. } \thispagestyle{empty} \newpage
\renewcommand
{\thesection}{\arabic{section}}

\tolerance=10000

\section{Introduction}

Gutzwiller's trace formula of 1971 expresses the density of states $g(E)$ of
a quantum mechanical system approximately as a sum over all periodic
classical orbits \cite{1}. Later, Balian and Bloch \cite{2} presented a
formula which also applies to nonintegrable systems. It
arose from a study of sound spectra in cavities with reflecting walls
of arbitrary shape in two and more dimensions. Gutzwiller's formula applies
only to systems with isolated orbits. It fails if there exist degenerate
families of periodic orbits connected  by continuous
symmetries \cite{3}.
The problem arise in the derivation
of Gutzwiller's formula from a stationary phase approximation to the trace
integral over the semiclassical Green function at fixed energy. It contains
an oscillating exponential of the eikonal function $S({\bf x},{\bf x}%
;E)=\oint {\bf p}\cdot d{\bf x}$ of the periodic orbits passing through the point $%
{\bf x}$. A continuous symmetry makes this independent of ${\bf x}$ over an
entire spatial region swept out by the symmetry operations. Then the second
derivatives of the eikonal function vanishes in that region, resulting in a
divergence of the stationary-phase integral. Strutinsky and coworkers \cite
{4} removed these divergences by going back the convolution integral in the
time-dependent propagator and performing exactly as many integrations in
that integral and in the trace integral over the Green function, as there
are independent parameters describing  the degeneracy.
Later, Creagh and Littlejohn \cite{5}
pursued the same idea in a generalized phase space which also contains room
for the continuous symmetry of the system. For integrable systems, their
procedure is similar to that of Berry and Tabor \cite{6} who derived a trace
formula for integrable systems employing the action-angle variables.

All this development has so far been restricted to the {\em nonrelativistic\/%
} regime where the particle solves the Schr\"odinger equation in some
external time-independent potential. The purpose of this paper is to
begin adapting
the methods to {\em relativistic\/} particles described by the Klein-Gordon
equation in the external potential. Our final result will be a relativistic
generalization of Gutzwiller's trace formula, expressing the density of
states as a topological sum over the relativistic closed classical orbits.
The formula is applicable to integrable relativistic classical systems.

Relativistic quantum mechanics is of course not really a
consistent theory. At relativistic velocities, particles will be created and
absorbed, and the particle number is no longer conserved, thus violating the
current conservation law of the Klein-Gordon equation. Quantum field theory
is certainly the appropriate tool to describe relativistic particles. In the
classical regime, however, the particle number is fixed
and these problems are absent, so that a
semiclassical expression for the density of states in terms of relativistic
classical periodic orbits is a consistent
approximation expected to render a reliable results
for those
systems in which particle creation and annihilation play only a minor role.

\section{Relativistic Quantum-Mechanical Trace Formula}

Consider a relativistic particle of mass $m$ in an external time-independent
potential $V({\bf x})$, whose quantum mechanics is governed by the
Klein-Gordon equation
\begin{equation}
\left\{ \lbrack i\hbar \partial _{t}+mc^{2}-V({\bf x})]^{2}-c^{2}\hbar
^{2}\left( \partial _{\mbox{{\scriptsize$\bf{x}$}}}-i\frac{e}{c\hbar }{\bf A}%
\right) ^{2}-m^{2}c^{4}\right\} \phi ({\bf x},t)=0.  \label{1.1}
\end{equation}
where $c$ and $\hbar $ are speed of light and Planck's constant, and ${\bf A}%
({\bf x})$ is a magnetic vector potential. We have shifted the energy origin
to the rest energy $mc^{2}$ in order to have a smooth limit to
nonrelativistic bound-state energies. Since the potentials are
time-independent, the wave functions can be factorized as $\phi ({\bf x}%
,t)=e^{-iEt/\hbar }\Psi ({\bf x})$, and (\ref{1.1}) takes the
Schr\"odinger-like form
\begin{equation}
\hat{{\cal H}}_{E}\Psi ({\bf x})={\varepsilon }\Psi ({\bf x}),  \label{1.2}
\end{equation}
where
\begin{equation}
{\varepsilon }\equiv \frac{E^{2}-m^{2}c^{4}}{2mc^{2}},  \label{1.3}
\end{equation}
and $\hat{{\cal H}}_{E}$ is the Hamilton operator
\begin{equation}
\hat{{\cal H}}_{E}=\hat{{\bf p}}^{2}/2m+[2EV({\bf x})-V^{2}({\bf x}%
)]/2mc^{2},  \label{1.4}
\end{equation}
with ${\bf p}=-i\hbar \partial _{\mbox{{\scriptsize$\bf{x}$}}}$. It is
useful to view (\ref{1.2}) as a special case of a more general eigenvalue
equation
\begin{equation}
\hat{{\cal H}}_{E}\Psi ({\bf x})={\cal E}\Psi ({\bf x}),  \label{1.5}
\end{equation}
which arises from a Schr\"odinger-like equation
\begin{equation}
\hat{{\cal H}}_{E}\Psi ({\bf x},\tau )=i\hbar \partial _{\tau }\Psi ({\bf x}%
,\tau )  \label{1.6}
\end{equation}
by a factorized ansatz $\Psi ({\bf x},\tau )=e^{-i{\cal E}\tau /\hbar }\Psi (%
{\bf x})$. Then the variable $\tau $ plays the role of a pseudotime, and the
Hamilton operator $\hat{{\cal H}}_{E}$ is the pseudotime-evolution operator
governing the $\tau $-dependence of the system.

Let $\Psi _{\mbox{{\scriptsize$\bf{n}$}}}({\bf x})$ be the eigenfunctions of
Eq.~(\ref{1.5}) with eigenvalues ${\cal E}_{E}({\bf n})$. Then
the physical energies
$E_{\sbf n}$ of the particle are given by those values of $E$ at which the
pseudoenergy is equal to $\varepsilon$:
\begin{equation}
  {\cal E}_{E_{\sbf n}}({\bf n})=\varepsilon.
\label{@12}\end{equation}

As an example, consider the Coulomb potential $V(r)=-e^{2}/r$ of the
relativistic hydrogen atom. Equation (\ref{1.5}) leads to the radial
eigenvalue equation
\begin{equation}
\frac{d^{2}R(r)}{dr^{2}}+\frac{2}{r}\frac{dR(r)}{dr}+\left[ \frac{2m}{\hbar
^{2}}\left( {\cal E}_{E}+\frac{Ee^{2}/mc^{2}}{r}\right) -\frac{l(l+1)-\alpha
^{2}}{r^{2}}\right] R(r)=0.  \label{1.10}
\end{equation}
Its solutions yield the bound state pseudoenergies depending on principal
quantum number $n$ and angular momentum $l$, but degenerate in the azimuthal
quantum number $m$:
\begin{equation}
{\cal E}_{E}(n,l,m)=-\frac{E^{2}/mc^{2}}{2}\frac{\alpha ^{2}}{\left[ \left(
n-l-1/2\right) +\sqrt{\left( l+1/2\right) ^{2}-\alpha ^{2}}\right] ^{2}}%
,\quad \left\{
\begin{array}{l}
n=1,2,3,\cdots  \\
l=0,1,2,\cdots
\end{array}
\right. .  \label{1.11}
\end{equation}
Inserting these into Eq. (\ref{@12}), we obtain
the well-known relativistic bound energies of the Coulomb system:
\begin{equation}
E_{n,l}=\pm mc^{2}\left[ 1+\frac{\alpha ^{2}}{\left[ \left( n-l-1/2\right) +%
\sqrt{\left( l+1/2\right) ^{2}-\alpha ^{2}}\right] ^{2}}\right] ^{-1/2}.
\label{1.12}
\end{equation}

The complete information on the spectrum of eigenvalues
of the Klein-Gordon equation (\ref{1.1}) is contained in the pole
terms of the trace of the resolvent $\hat{R}(E)\equiv i({{\varepsilon }-\hat{%
{\cal H}}_{E}({\bf n})+i\eta })^{-1}$:
\begin{equation}
r(E)\equiv {\rm Tr}\,\hat{R}(E)=\,i\,{\rm Tr}\,[{{\varepsilon }-\hat{{\cal H}%
}_{E}({\bf n})}+i\eta ]^{-1},  \label{1.7}
\end{equation}
where the infinitesimal positive quantity $\eta $ guarantees the causality
of the time dependence of the Fourier transform of $r(E)$. The imaginary
part of $r(E)$ defines the {\em density of states\/}:
\begin{equation}
g(E)=\frac{1}{\pi }{\rm Im}\,r(E)={\rm Tr}\,\delta ({\varepsilon }-\hat{%
{\cal H}}_{E}).  \label{1.8}
\end{equation}
In terms of the eigenvalues
${\cal E}_{E}({\bf n})$, the density (\ref{1.8})
has the spectral representation
\begin{equation}
g(E)=\sum_{\mbox{{\scriptsize$\bf{n}$}}}\delta ({{\varepsilon }-{\cal E}_{E}(%
{\bf n})}),  \label{1.9}
\end{equation}
where the sum over ${\bf n}$ covers all quantum numbers.
This sum will now be performed in a semiclassical approximation as a sum over
periodic classical orbits.

For the sake of generality, we assume that the particle moves in $D$%
-dimensions, and assume that the motion has been transformed to $D$ cyclic
coordinates whose motion can easily be quantized ({\em torus quantization\/}%
). The labels ${\bf n}$
will then be integer-valued vectors ${\bf n}=(n_{1},n_{2},\cdots ,n_{D})$
with non-negative components $n_{i}$. For the purpose of deriving a
semiclassical approximation to (\ref{1.9}), we convert each sum over $%
n_{i}=0,1,2,\dots $
in Eq. (\ref{1.9})
 into an integral with the help of the Poisson summation
formula \cite{PI,7}
\begin{equation}
\sum_{n=0}^{\infty }f(n)=\sum_{k=-\infty }^{\infty }\int_{0^{-}}^{\infty
}f(n)e^{2\pi ikn}dn . \label{1.13}
\end{equation}
 Here we have assumed that the function $f(n)$ and its
derivatives with respect to $n$ vanish at infinity, and the lower limit $%
0^{-}$ on the integral sign indicates that the integration starts
on the left-hand side
of the origin to include the entire $\delta $-function generated by the sum
over $k$. The superscript will be omitted in the sequel. Thus we obtain
\begin{equation}
g(E)=\sum_{{\mbox{{\scriptsize$\bf{k}$}}}}\int d^{D}n\,\delta \left( {%
\varepsilon -}{\cal E}_{E}({\bf n})\right) e^{2\pi i{\bf k}{\bf n}},
\label{1.14}
\end{equation}
where each component of the integer-valued vector ${\bf k}%
=(k_{1},k_{2},\cdots ,k_{D})$ runs form minus to plus infinity, while
the now continuous
variables
$n_{i}$ are integrated from $0^{-}$ to infinity.

For integrable systems, the integration variables $n_{i}$ in Eq. (\ref{1.14}%
) can be replaced by the values of the action integrals appearing in the
relativistic quantum conditions \cite{8}
\begin{equation}
I_{i}=\frac{1}{2\pi }\oint_{C_{i}}{\bf p}\cdot d{\bf x}=\left( n_{i}+\frac{%
\mu _{i}}{4}\right) \hbar ,  \label{1.15}
\end{equation}
where ${\bf p}$ is the relativistic momentum of the point particle along
closed loops $C_{i}$ on an invariant torus. The quantum numbers $n_{i}$ are
the same nonnegative integers as above, while $\mu _{i}$ are the numbers of
conjugate points along the orbit $C_{i}$. Thus we can rewrite Eq.~(\ref{1.14}%
) as
\begin{equation}
g(E)=\frac{1}{\hbar ^{D}}\sum_{{\bf k}}e^{-i{\bf k\cdot {{%
\mbox{\egtmbf\symbol{'026}}}}}\,\pi /2}\int_{\hbar {\mbox{\egtmbf%
\symbol{'026}}}_{1}/4}^{\infty }dI_{1}\int_{\hbar {\mbox{\egtmbf%
\symbol{'026}}}_{2}/4}^{\infty }dI_{2}\cdots \int_{\hbar {%
\mbox{\egtmbf\symbol{'026}}}_{D}/4}^{\infty }dI_{D}\,\delta \left( {%
\varepsilon -}{\cal E}_{E}({\bf I})\right) e^{2\pi i{\bf k\cdot I}},
\label{1.16}
\end{equation}
where we have changed the argument of ${\cal E}_{E}({\bf n})$ to ${\cal E}%
_{E}({\bf I})$, and introduced vectors ${\mbox{\twlmbf\symbol{'026}}}%
=(\mu _{1},\mu _{2},\dots ,\mu _{D})$.

Consider now the lowest term with ${\bf k}=0$, for which the oscillating
exponentials in Eq. (\ref{1.16}) are absent. It contributes a smooth density
of states
\begin{equation}
\bar{g}(E)=\frac{1}{\hbar ^{D}}\int_{0}^{\infty }dI_{1}\int_{0}^{\infty
}dI_{2}\cdots \int_{0}^{\infty }dI_{D}\,\delta \left( {\varepsilon -}{\cal E}%
_{E}({\bf I})\right) ,  \label{1.17}
\end{equation}
where the lower bounds of the integral has been moved to zero, since the
classical orbits for ${\bf k}=0$ have zero length, making ${%
\mbox{\twlmbf\symbol{'026}}}$ equal to zero. The multiple integral (\ref
{1.17}) is just the classical density of states
\begin{equation}
\bar{g}_{{\rm cl}}(E)=\frac{1}{(2\pi \hbar )^{D}}\int \int d^{D}{p}\,d^{D}{q}%
\,\delta \left( {\varepsilon -}{\cal E}_{E}({\bf p},{\bf q})\right) ,
\label{1.18}
\end{equation}
which in cyclic coordinates reads
\begin{equation}
\bar{g}_{{\rm cl}}(E)=\frac{1}{(2\pi \hbar )^{D}}\int_{0}^{\infty
}dI_{1}\int_{0}^{2\pi }d\varphi _{1}\int_{0}^{\infty }dI_{2}\int_{0}^{2\pi
}d\varphi _{2}\cdots \int_{0}^{\infty }dI_{D}\int_{0}^{2\pi }d\varphi
_{D}\,\delta \left( {\varepsilon -}{\cal E}_{E}({\bf I})\right) ,
\label{1.19}
\end{equation}
reducing to (\ref{1.17}) after integrating out the angular variables. The
classical density of states is also referred to as the {\em Thomas-Fermi density\/}
\cite{PI2},

We now turn to the oscillating ${\bf k}\neq {\bf 0}$ parts of $g(E)$. With
the help of the integral representation for the $\delta $-function
\begin{equation}
\delta \left( {\varepsilon -}{\cal E}_{E}({\bf I})\right) =\frac{1}{2\pi
\hbar }\int_{-\infty }^{\infty }d\tau \,e^{i\tau \left[ {\varepsilon -}{\cal %
E}_{E}({\mbox{{\scriptsize$\bf{I}$}}})\right] /\hbar },  \label{1.20}
\end{equation}
we rewrite this as
\begin{equation}
\delta g(E)=\frac{1}{2\pi \hbar }\int_{-\infty }^{\infty }d\tau \frac{1}{%
\hbar ^{D}}\sum_{{\mbox{{\scriptsize$\bf{k}$}}}}{}^{\!\prime }e^{-i{\bf %
k\cdot {\mbox{\egtmbf\symbol{'026}}}}\,\pi /2}\int_{\hbar {%
\mbox{\egtmbf\symbol{'026}}}_{1}/4}^{\infty }dI_{1}\int_{\hbar {%
\mbox{\egtmbf\symbol{'026}}}_{2}/4}^{\infty }dI_{2}\cdots \int_{\hbar {%
\mbox{\egtmbf\symbol{'026}}}_{D}/4}^{\infty }dI_{D}e^{{i}\left\{ 2\pi {\bf %
k\cdot I+}\tau [{\varepsilon -}{\cal E}_{E}({\bf I})]\right\} /\hbar },
\label{1.21}
\end{equation}
where the primes on the summation symbols indicate the omission of ${\bf k}=%
{\bf 0}$. The integrals over $I_{i}$ and $\tau $ are now evaluated in the
stationary phase approximation. Let us abbreviate the action in the exponent
by
\begin{equation}
A_{{\bf k}}({\bf I},\tau )=2\pi {\bf k\cdot I+}\tau [{\varepsilon -}{\cal E}%
_{E}({\bf I})].  \label{1.22}
\end{equation}
Its extrema lie at some ${\bf I}=\bar{{\bf I}}$, $\tau =\bar{\tau}$, where
\begin{equation}
\left. \frac{\partial A_{{\bf k}}}{\partial I_{i}}\right| _{{\bf I}=\bar{%
{\bf I}},\tau =\bar{\tau}}=0,~~~~~\left. \frac{\partial A_{{\bf k}}}{%
\partial \tau }\right| _{{\bf I}=\bar{{\bf I}},\tau =\bar{\tau}}=0.~~~~~
\label{1.23}
\end{equation}
The first set of equations yields the semiclassical quantization condition
\begin{equation}
2\pi k_{i}=\bar{\tau}\omega _{i}(\bar{{\bf I}}),\quad i=1,2,\cdots ,D,
\label{1.24}
\end{equation}
where
\begin{equation}
\omega _{i}({\bf I})\equiv \frac{\partial {\cal E}_{E}({\bf I})}{\partial I_{i}}
 \label{@}\end{equation}
at
$\bar{{\bf I}}$ are the angular velocities for the pseudoenergy ${%
\varepsilon }$. The solutions of Eq. (\ref{1.24}) yield actions $\bar{{\bf I}%
}$ as nonlinear functions of ${\bf k}$ and $\bar{\tau}$:
\begin{equation}
\bar{{\bf I}}=\bar{{\bf I}}({\bf k},\bar{\tau}).  \label{1.25}
\end{equation}
From Eq. (\ref{1.24}) we obtain the important relation for the resonant tori
\begin{equation}
\frac{k_{i}}{k_{j}}=\frac{\omega _{i}}{\omega _{j}},\quad i,j=1,2,\cdots ,D.
\label{1.26}
\end{equation}
Since ${ k_i}$ ere integer numbers, the orbits on the torus must
have commensurate frequencies, so that only closed periodic orbits
contribute to the density of states in the saddle point approximation. This
establishes the connection between $\delta g(E)$ and the relativistic
periodic orbits of the classical system. If the frequencies are not
commensurate, the orbits do not close although the motion is still confined
to the torus. Such orbits are called multiply periodic or quasi-periodic.

Each relativistic periodic orbit is specified by ${\bf k}$; it closes after $%
k_{1}$ turns by $2\pi $ of the angle $\varphi _{1},$ $k_{2}$ turns of $%
\varphi _{2}$, \dots ~.
Thus
${\bf k}$ plays the role of
an {\em index vector\/} characterizing the
topology of the periodic orbits.
For this reason, the sums in Eq. (\ref{1.16}) is also called topological sum.
Note that Eq. (\ref{1.26}) admits only $k_{i}$-values of the same sign.

The second equation in (\ref{1.23}) specifies $\bar{\tau}$ via
\begin{equation}
{\varepsilon }-{\cal E}_{E}\left( {\bf \bar{I}(}{\bar{\tau}}(E))\right) =0.
\label{1.27}
\end{equation}
Having determined the saddle points, the semiclassical approximation
requires the calculation of the effect of the quadratic fluctuations around
these. For this we expand Eq. (\ref{1.22}) up to the quadratic terms, and
shift the integration variables from ${\bf I}$ to ${\bf I}^{\prime }\equiv
{\bf I}-\bar{{\bf I}}$. The lower bound of the integrals is then transformed
into $\hbar {\mbox{\twlmbf\symbol{'026}}}/4-\bar{{\bf I}}$. For sufficiently
large actions $\bar{I}_{i}$, the sharpness of the extrema at small $\hbar $
allows us to move the lower bounds to minus-infinity. This approximation is
excellent for highly excited states. We now perform the Gaussian integrals
and obtain the oscillating part of the relativistic density of states
\begin{equation}
\delta g^{(2)}(E)=\frac{1}{2\pi }\sqrt{2\pi /\hbar }^{D+1}\sum_{%
\mbox{{\scriptsize$\bf{k}$}}}{}^{\!\prime }e^{-i{\bf k\cdot {%
\mbox{\egtmbf\symbol{'026}}}}\,\pi /2}e^{-i{\pi }\nu /4}\frac{1}{\bar{\tau}%
^{(D-1)/2}}\left| \det M\right| _{\bar{\mbox{{\scriptsize$\bf{I}$}}}%
}^{-1/2}e^{{i}2\pi {\bf k}\cdot {\bar{{\bf I}}}/\hbar },  \label{1.28}
\end{equation}
where $M$ is the stability matrix
\begin{equation}
M=\left(
\begin{array}{cc}
\bar{\tau}\displaystyle\frac{\partial ^{2}{\cal E}_{E}}{\partial
I_{i}\partial I_{j}} & \displaystyle \frac{\partial {\cal E}_{E}}{\partial
I_{i}} \\[2mm]
\displaystyle \frac{\partial {\cal E}_{E}}{\partial I_{j}} & 0
\end{array}
\right) .  \label{1.29}
\end{equation}
whose determinant is, according to formula
\begin{equation}
\det \left(
\begin{array}{ll}
A & B \\
C & D
\end{array}
\right) =\det A\,\det (D-C^{T}A^{-1}B)  \label{1.30}
\end{equation}
given by
\begin{equation}
\det M=\det H~\, {\mbox{\twlmbf\symbol{'041}}}^{T}\!H^{-1}{\mbox{\twlmbf%
\symbol{'041}}}.  \label{1.31}
\end{equation}
where
\begin{equation}
H_{ij}\equiv \frac{\partial ^{2}{\cal E}_{E}}{\partial I_{i}\partial I_{j}}.
\label{1.32}
\end{equation}
The Maslov index $\nu $ is equal to $N^{+}-N^{-}-N^{0}$, where $N^{\pm }$ denote
the numbers of positive and negative eigenvalues of matrix $H_{ij}$, and $%
N^{0}$ is unity (zero) if the sign of ${\mbox{\twlmbf\symbol{'041}}}%
^{T}H^{-1}{\mbox{\twlmbf\symbol{'041}}}$ is positive (negative). The second
factor in (\ref{1.30}) has been simplified using the equation of motion for
the cyclic variables ${\mbox{\twlmbf\symbol{'047}}}$:
\begin{equation}
\frac{d{{\mbox{\twlmbf\symbol{'047}}}}}{d\tau }=\nabla _{{\bf I}}{\cal E}%
_{E}({\bf I})={{\mbox{\twlmbf\symbol{'041}}}(I)},  \label{1.33}
\end{equation}
the right-hand side being also equal to
\begin{equation}
{\mbox{\twlmbf\symbol{'041}}}=\frac{2\pi {\bf k}}{\bar{\tau}}.  \label{1.34}
\end{equation}
Since for every ${\bf k}$ there is an equal contribution from $-{\bf k}$, we
may replace the exponential by a cosine and obtain
\begin{equation}
\delta g^{(2)}(E)=\frac{1}{2\pi }\sqrt{2\pi /\hbar }%
^{D+1}\sum_{\mbox{{\scriptsize$\bf{k}$}}}{}^{\!\prime }\frac{1}{\bar{\tau}%
^{(D-1)/2}}\left| \det H~{\mbox{\twlmbf\symbol{'041}}}^{T}H^{-1}{%
\mbox{\twlmbf\symbol{'041}}}\right| _{\bar{\mbox{{\scriptsize$\bf{I}$}}}%
}^{-1/2}\cos \left[ {{}2\pi {\bf k}\cdot \left( {\bar{{\bf I}}/\hbar }-{%
\mbox{\twlmbf\symbol{'026}}}/4\right) -\pi \nu /4}\right] .  \label{1.35}
\end{equation}
The relativistic trace formula (\ref{1.35}) gives us a basis for
understanding quantum phenomena at the relativistic level in terms of
classical orbits. In general, we just need to evaluate the classical ${\cal E%
}_{E}\left( {\bf \bar{I}}\right) $ for integrable systems, and consider some
shortest orbits. As in nonrelativistic systems, we expect astonishingly
accurate energy spectra from Eq. (\ref{1.35}).

\section{Three-Dimensional Relativistic Rectangular Billiard}

As a first application, consider the motion of a relativistic particle in a three-dimensional
rectangular billiard with sides of length $a_{1},a_{2},$ and $a_{3}$ along
$q_{1},q_{2},$ and $q_{3}$ axes. The quantum spectrum of Eq. (\ref{1.5})
with Dirichlet boundary condition is given by the pseudoenergies
\begin{equation}
{\cal E}_{E}({n_{1},n_{2},n_{3}})=\frac{\hbar ^{2}\pi ^{2}}{2m}\left( \frac{%
n_{1}^{2}}{a_{1}^{2}}+\frac{n_{2}^{2}}{a_{2}^{2}}+\frac{n_{3}^{2}}{a_{3}^{2}}%
\right) ,\quad n_{i}(i=1,2,3)=1,2,3,\cdots .  \label{2.1}
\end{equation}
The physical relativistic energy spectrum is obtained from Eq. (\ref{@12}):
\begin{equation}
E_{n_{1},n_{2},n_{3}}=\pm \sqrt{\pi ^{2}\hbar ^{2}c^{2}\left( \frac{n_{1}^{2}%
}{a_{1}^{2}}+\frac{n_{2}^{2}}{a_{2}^{2}}+\frac{n_{3}^{2}}{a_{3}^{2}}\right)
+m^{2}c^{4}.}  \label{2.2}
\end{equation}
As in the nonrelativistic case, this result is {\em exactly\/} reproduced by the relativistic quantization according
to Eq. (\ref{1.15}). The numbers $\mu _{i}$ are all equal to $4$,
since
the wave functions have Dirichlet boundary condition.
At every
every encounter with the wall,
the action  picks up a phase $\pi $.
The relativistic action variables are therefore
\begin{equation}
I_{i}=\frac{1}{2\pi }\oint p_{i}dq_{i}=n_{i}\hbar ,\quad i=1,2,3;\quad
n_{i}=1,2,3,\cdots .  \label{2.3}
\end{equation}
The classical Hamiltonian may be expressed as
\begin{equation}
{\cal E}_{E}({\bf I)=}\frac{\pi ^{2}}{2m}\left( \frac{I_{1}^{2}}{a_{1}^{2}}+%
\frac{I_{2}^{2}}{a_{2}^{2}}+\frac{I_{3}^{2}}{a_{3}^{2}}\right),  \label{2.4}
\end{equation}
and the corresponding angular frequencies are
\begin{equation}
\omega _{i}=\frac{\pi ^{2}}{ma_{i}^{2}}I_{i},\quad i=1,2,3.  \label{2.5}
\end{equation}
We now determine the saddle points $\bar{{\bf I}}.$ According to
Eq. (\ref{1.24}), these are given by
\begin{equation}
\left( \bar{I}_{1},\bar{I}_{2},\bar{I}_{3}\right) \left( \tau \right)
=\left( \frac{2ma_{1}^{2}k_{1}}{\tau \pi },\frac{2ma_{2}^{2}k_{2}}{\tau \pi }%
,\frac{2ma_{3}^{2}k_{3}}{\tau \pi }\right),  \label{2.6}
\end{equation}
leading to the pseudoenergies at the saddle point
\begin{equation}
{\cal E}_{E}({\bf \bar{I}(}\tau {\bf ))=}\frac{2m}{\tau ^{2}}%
\sum_{i=1}^{3}\left( a_{i}k_{i}\right) ^{2}.  \label{2.7}
\end{equation}
The saddle-point value of ${\tau }$ is determined by
(\ref{1.27}),
yielding
\begin{equation}
\bar{\tau}=\sqrt{\frac{2m}{{\varepsilon }}}\sqrt{\sum_{i=1}^{3}\left(
a_{i}k_{i}\right) ^{2}}.  \label{2.8}
\end{equation}
From these saddle point values, we obtain
\begin{equation}
\left. \sum_{ij}\omega _{i}H_{ij}^{-1}\omega _{j}\right| _{\bar{%
\mbox{{\scriptsize$\bf{I}$}}}(\bar{\tau})}=2{\varepsilon }  \label{2.9}
,\end{equation}
so that the sign of $\omega _{i}H_{ij}^{-1}\omega _{j}$ is positive
and the number  $N^0$ in the Maslov index $ \nu =N^{+}-N^{-}-N^{0}$
 vanishes. The
determinant of the second-derivative matrix is
\begin{equation}
\det \frac{\partial ^{2}{\cal E}_{E}}{\partial I_{i}\partial I_{j}}=\frac{%
\pi ^{6}}{m^{3}\,a_{1}^{2}a_{2}^{2}a_{3}^{2}}.  \label{2.10}
\end{equation}
All eigenvalues of the matrix
$\lfrac{\partial ^{2}{\cal E}_{E}}{\partial I_{i}\partial I_{j}}$ are positive.
Thus
we identify the
indices $N^{+}=3,~N^{-}=0$. Inserted into Eq.~(\ref{1.35}), we
finally obtain for the oscillating part of the relativistic density of states
\begin{equation}
\delta g^{(2)}(E)=\frac{%
\pi }{4E_{0}}\sqrt{\frac{{\varepsilon }}{E_{0}}} \frac{a_{1}a_{2}a_{3}}
{L^{3}}
\sum_{k_{1},k_{2},k_{3}=-\infty }^{\infty}{}^{\!\!\!\!\!\!\!\!\!\!\!\!\!\!
 \prime} ~~~~~j_{0}\left( \frac{S\left( {\bf k}\right) }{\hbar }\right) ,
\label{2.11}
\end{equation}
where $j_{0}(x)$ is the spherical Bessel function of order zero
$j_{0}(x)=\sin (x)/x.$ The symbol  $L$
denotes some lenght scale which may be any average of the three length scales $%
a_1,\,a_{2}$,or $a_{3}$,
while
\begin{equation}
E_{0}\equiv \frac{\pi ^{2}\hbar ^{2}}{2mL^{2}}  \label{2.12}
\end{equation}
denotes the energy associated with
$L$.
The quantity $S\left( {\bf k}\right) $ is
\begin{equation}
S\left( {\bf k}\right) =\frac{1}{c}\sqrt{%
E^{2}-m^{2}c^{4}}~\,2\sqrt{%
k_{1}^{2}a_{1}^{2}+k_{2}^{2}a_{2}^{2}+k_{3}^{2}a_{3}^{2}}=p~L_{\sbf k}.  \label{2.13}
\end{equation}
It is precisely the relativistic eikonal
$p\,L_{\sbf k}$ of the classical periodic
orbits of momentum $p$
and total length $L_{\sbf k}$.
In general, the inclusion of
only a few shortest orbits in Eq. (\ref{2.11}) yields the correct positions
of the quantum energy levels. The three-dimensional relativistic rectangular
billiard may serve as a prototype of the relativistic semiclassical
treatment for arbitrary billiard systems.

Let us compare the calculation of (\ref{2.11})
from our relativistic trace formula (\ref{1.35})
with a direct calculation
from an inverse
Laplace transformation of partition function $Z(\beta )$, i.e,
\begin{equation}
g(E)=\frac{1}{2\pi i}\int_{\epsilon -i\infty }^{\epsilon +i\infty }d\beta \,
e^{\beta {\varepsilon }}Z(\beta ),  \label{2.14}
\end{equation}
where the partition function is given by
\begin{equation}
Z(\beta )=\sum_{n_{1}=1}^{\infty }\sum_{n_{2}=1}^{\infty
}\sum_{n_{3}=1}^{\infty }\exp \left\{ -\beta {\cal E}_{E}(n_{1},n_{2},n_{3})%
\right\} ,  \label{2.15}
\end{equation}
with the pseudoenergies ${\cal E}_{E}({n_{1},n_{2},n_{3}})$ of
Eq.~(\ref{2.1}). The problem is the same
as in the calculation of the Casimir energy for
the box. Since (\ref{2.15}) is a product of three independent sums
\begin{equation}
Z_{i}(\beta )=\sum_{n_{i}=1}^{\infty }\exp \left\{ -\beta {\cal E}_{E}({n_{i}%
})\right\} ,\quad i=1,2,3,  \label{2.16}
\end{equation}
we may process each sum separately. Applying the Poisson formula (\ref{1.13})
to the sum over $n_i$ we find
\[
Z_{i}(\beta )=\sum_{k_{i}=-\infty }^{\infty }\int_{-\infty }^{\infty
}dn_ie^{-\beta E_{0}n_i^{2}L^{2}/a_{i}^{2}}e^{2\pi ik_{i}n_i}-\frac{1}{2}
~~~~~~~~~~~~\]
\begin{equation}
\!\!\!\!\!\!\!\!\!\!\!=\frac{1}{2}\frac{a_{i}}{L}\sqrt{\frac{\pi }{\beta E_{0}}}%
\sum_{k_{i}=-\infty }^{\infty }e^{-\left( \pi ma_{i}\right) ^{2}/\beta
E_{0}L^{2}}-\frac{1}{2}.  \label{2.17}
\end{equation}
Inserting this into (\ref{2.15}), and using the integral formula
\begin{equation}
\frac{1}{2\pi i}\int_{\epsilon -i\infty }^{\epsilon +i\infty }\frac{d\beta }{%
\beta ^{\mu +1}}e^{\beta {\varepsilon }}e^{-\kappa /\beta }=\left( \frac{{%
\varepsilon }}{\kappa }\right) ^{\mu /2}J_{\mu }\left( 2\sqrt{\kappa {%
\varepsilon }}\right) ,  \label{2.18}
\end{equation}
we obtain the exact level density of the
relativistic three-dimensional rectangular box
\begin{equation}
g(E)=g^{(3)}(E)-\frac{1}{2}\left[
g_{12}^{(2)}(E)+g_{23}^{(2)}(E)+g_{31}^{(2)}(E)\right] +\left[
g_{1}^{(1)}(E)+g_{2}^{(1)}(E)+g_{3}^{(1)}(E)\right] -\frac{1}{8}\delta
\left( {\varepsilon }\right) .  \label{2.19}
\end{equation}
The leading term comes from a proper three-fold sum, and is given by
\begin{equation}
\delta g^{(2)}(E)=\sum_{k_{1},k_{2},k_{3}=-\infty }^{\infty }\frac{\pi }{%
4E_{0}}\sqrt{\frac{{\varepsilon }}{E_{0}}}\left( \frac{a_{1}a_{2}a_{3}}{L^{3}%
}\right) j_{0}\left( \frac{S\left( {\bf k}\right) }{\hbar }\right)
\label{2.20}
\end{equation}
with $S\left( {\bf k}\right) $ of Eq.~(\ref{2.13}). This agrees
with the
semiclassical result (\ref{2.11}). The second set of
terms gives corrections from the faces of the box:
\begin{equation}
g_{ij}^{(2)}(E)=\frac{\pi }{4}\frac{1}{E_{0}}\frac{a_{i}a_{j}}{L^{2}}%
\sum_{k_{1},k_{2}=-\infty }^{\infty }J_{0}\left( \frac{S_2\left(
k_{1},k_{2}\right) }{\hbar }\right) ,  \label{2.21}
\end{equation}
where
\begin{equation}
S_2\left( k_{1},k_{2}\right)=\frac{1}{c}\sqrt{%
E^{2}-m^{2}c^{4}}~~2\sqrt{k_{1}^{2}a_{i}^{2}+k_{2}^{2}a_{j}^{2}}  =p\,L_{k_{1},k_{2}} \label{2.22}
\end{equation}
are the eikonals
of the orbits on the faces.
The functions $g_{ij}^{(2)}(E)$ are
the level densities of the planar facial ``boxes''.

The third set of
terms in (\ref{2.19})
stems from the edges of the box, being
the level density for these one-dimensional
``boxes'' of length $a_{i}$:
\begin{equation}
g_{i}^{(1)}(E)=\frac{a_{i}}{2L\sqrt{E_{0}{\varepsilon }}}\sum_{k_i=-\infty
}^{\infty }\cos \left( \frac{S_1\left( k_i\right) }{\hbar }\right) ,
\label{2.23}
\end{equation}
where
\begin{equation}
S_1\left( k\right) =\frac{1}{c}\sqrt{E^{2}-m^{2}c^{4}}~2ka_{i}=p\,L_{k_{1},k_{2}} .  \label{2.24}
\end{equation}
These boundary terms can be obtained also from the general trace formula (%
\ref{1.13}) by calculating higher-order corrections to the semiclassical
approximation (\ref{1.35}).

The last term in (\ref{2.19}) is a delta function at ${\varepsilon}=0$ which
does not contribute to the level density at ${\varepsilon}>0$.

The
 classical (Thomas-Fermi) contribution to the density of states is
\begin{equation}
\bar{g}(E)=\frac{1}{E_{0}}\left( \frac{\pi }{4}\sqrt{\frac{{\varepsilon}}{%
E_{0}}}\frac{{V}_3}{L^{3}}-\frac{\pi }{8}\frac{{V_2}}{2L^{2}}+\frac{1}{8}%
\sqrt{\frac{E_{0}}{{\varepsilon}}}\frac{{V_1}}{L}\right) .  \label{2.25}
\end{equation}
Here ${V}_3=a_{1}a_{2}a_{3}$ is the volume of the box, ${V_2=}2\left(
a_{1}a_{2}+a_{2}a_{3}+a_{1}a_{3}\right) $ the total surface, and $%
V_1=a_{1}+a_{2}+a_{3})$  the sum of the edge lengths.

\section{Concluding remark}

For relativistic integrable systems, we have derived a semiclassical trace
formula by transforming the relativistic quantization conditions into the
topological sum involving all closed relativistic classical orbits.
Certainly, our final result (\ref{1.35}) can
also be obtained by an
{\sf ab initio} procedure,
 starting out from the relativistic path integral for
the relativistic fixed-energy amplitude representation \cite{PI,9,10}
\begin{equation}
G({\bf {x}}_{b},{\bf {x}}_{a};E)=\frac{\hbar }{2Mc}\int_{0}^{\infty }dL\int
D\rho \Phi \left[ \rho \right] \int D^{D}xe^{iA_{E}/\hbar },  \label{3.1}
\end{equation}
with the action
\begin{equation}
A_{E}\left[ x,\dot{x}\right] =\int_{\tau _{a}}^{\tau _{b}}d\tau \left[ \frac{%
M}{2\rho \left( \tau \right) }{\bf \dot{x}}^{2}\left( \tau \right) +{\frac{e%
}{c}{\bf A\cdot \dot{x}}(\tau })+\frac{\rho (\tau )}{2Mc^{2}}\left( E-V({\bf %
x})\right) ^{2}-\rho \left( \tau \right) \frac{Mc^{2}}{2}\right],
\label{3.2}
\end{equation}
where $L$ is defined by
\begin{equation}
L=\int_{\tau _{a}}^{\tau _{b}}d\tau \rho (\tau ),  \label{3.3}
\end{equation}
with $\rho (\tau )$ being an arbitrary dimensionless fluctuating scale
variable, and $\Phi [\rho ]$ is some convenient gauge-fixing functional,
such as $\Phi \left[ \rho \right] =\delta \left[ \rho -1\right] $.
The prefactor $\hbar /Mc$ in (\ref{3.1}) is the Compton
wave length of a particle of mass $M$, the field ${\bf A(x)}$ is
the vector potential,
$V({\bf x})$ the scalar potential, $E$ the system energy, and ${\bf x}$
 the spatial part of the $D+1$ -dimensional vector $x=({\bf x},i\tau )$. This path
integral forms the basis for studying relativistic potential problems.
Choosing $\rho (\tau )$ to be equal to unity,
the amplitude
(\ref{3.1})
becomes
\begin{equation}
G({\bf {x}}_{b},{\bf {x}}_{a};E)=\frac{\hbar }{2Mc}\int_{0}^{\infty }dL\exp
\left[ \frac{i}{\hbar }{\varepsilon }L\right] \int {\cal D}^{D}x\exp \left[
\frac{i}{\hbar }A_{E}\right] ,  \label{3.4}
\end{equation}
where the fixed-energy action $A_{E}$ is given by
\begin{equation}
A_{E}=\int_{0}^{L}d\tau \left\{ \frac{M}{2}{\bf \dot{x}}^{2}\left( \tau
\right) +{\frac{e}{c}{\bf A\cdot \dot{x}}(\tau })+\frac{1}{2Mc^{2}}\left[ V^2(%
{\bf x})-2EV({\bf x})\right] \right\} .  \label{3.5}
\end{equation}
The semiclassical
approximation to the relativistic
fixed-energy amplitude (\ref{3.4}) is
\cite{11}
\[
\!\!\!\!\!\!\!\!\!\!\!\!\!\!\!\!\!\!\!G_{{\rm sc}}({\bf {x}}_{b},{\bf {x}}%
_{a};E)=\frac{\hbar }{2Mc}\frac{1}{(2\pi \hbar i)^{D/2}}\sum_{{\rm %
class.traj.}}\int_{0}^{\infty }dL
\,e^{i\varepsilon L/\hbar }
\]
\begin{equation}
\times \det \left[ -\partial _{x_{b}^{i}}\partial _{x_{a}^{j}}A_{E}({\bf {x}}%
_{b},{\bf {x}}_{a};L)\right] ^{1/2}e^{ {i}A_{E}({\bf {x}}%
_{b},{\bf {x}}_{a};L)/\hbar -i\pi \nu /2 }.  \label{3.6}
\end{equation}
The associated density of states is obtained from the trace of (\ref{3.6}%
):
\[ \!\!\!\!\!\!\!\!\!\!\!\!\!\!\!\!\!\!
\int d^{D}x\,G_{{\rm sc}}({\bf {x}}_{b},{\bf {x}}_{a};E)=\frac{\hbar }{2Mc}%
\frac{1}{(2\pi \hbar i)^{D/2}}\sum_{{\rm class.traj.}}\int_{0}^{\infty
}dL\,e^{{i}{\varepsilon }L/\hbar  }~~~~~~~~~~~~~
\]
\begin{equation}
\times \int d^{D}x\det \left[ -\partial _{x_{b}^{i}}\partial
_{x_{a}^{j}}A_{E}({\bf {x}}_{b},{\bf {x}}_{a};L)\right] ^{1/2}e^{
{i}A_{E}({\bf {x}}_{b},{\bf {x}}_{a};L)/\hbar -i\pi \nu /2 }.
\label{3.7}
\end{equation}
The trace operation in Eq. (\ref{3.6}) is integration over all periodic
orbits in the pseudotime ``$L"$. If the relativistic systems is integrable,
it can be expressed in terms of action-angle variables as
\[
\int d^{D}x\det \left[ -\partial _{x_{b}^{i}}\partial _{x_{a}^{j}}A_{E}({\bf
{x}}_{b},{\bf {x}}_{a};L)\right] ^{1/2}e^{iA_{E}({\bf {%
x}}_{b},{\bf {x}}_{a};L)/\hbar -i\pi \nu /2}~~~~~~~~~~~~~~~~~~~~~~~~
\]
\begin{equation}
=\sum_{{\bf k}}\int_{0}^{2\pi }d^{D}\varphi L^{-D/2}\det \left[ \frac{%
\partial ^{2}{\cal E}_{E}\left( {\bf I}\right) }{\partial I_{i}\partial I_{j}%
}\right] ^{-1/2}e^{i\left[ 2\pi {\bf I\cdot k}-{\cal E}_{E}({\bf I})L\right]
/\hbar -i\pi \nu /2}  \label{3.8}
\end{equation}
\begin{equation}
A_{E}({\bf {x}}_{b},{\bf {x}}_{a};L)={\bf I}\cdot \left( {\ {%
\mbox{\twlmbf\symbol{'047}}}}_{b}-{\ {{\mbox{\twlmbf\symbol{'047}}}}}%
_{a}\right) -{\cal E}_{E}({\bf I})L=2\pi {\bf I\cdot k}-{\cal E}_{E}({\bf I}%
)L,  \label{3.9}
\end{equation}
thus establishing contact with the earlier treatment
in which $\tau $ plays the role of $L$.
~\newline
\centerline{ACKNOWLEDGMENTS} \newline
This work was supported by the National Youth Council of the ROC under
contract number NYC300375.

\end{document}